\documentclass[12pt]{article}
\usepackage[latin1]{inputenc}
\begin{document}

\title{UM ERRO COM UM SÉCULO}

\author{António Brotas\footnote{brotas@fisica.ist.utl.pt}}

\maketitle

\begin{abstract}

The \textit{rigid bodies}  must be in Relativity the \textit{deformable bodies}  where the longitudinal waves propagate 
with the  maximun speed  $c$. 

In 1909, Born studied the  \textit{relativistic underformable body} but made the mistake of calling it  \textit{rigid}.
The \textit{rigid body}  one can find in all Relativity books is, in fact,   Born's  \textit{undeformable body} of 1909. 
This error was at the origin of a lot of difficulties and paradoxes not yet clarifed in any relativity book.  

The \textit{relativistic elastic laws for rigid bodies in one dimension} were discovered by Mc Crea, in 1952,
 and by  A.Brotas, in 1968. 
The generalization of these laws for 2 and 3 dimensions resolve all those paradoxes. 
We propose the introdution of these laws in the elementary courses of Relativity.

\end{abstract}

Na realidade há corpos mais ou menos rígidos consoante as deformações que sofrem quando sujeitos a forças. Acontece
 que a maioria dos sólidos que encontramos têm nas situações correntes deformações tão pequenas que as não notamos. 
A observação destes sólidos permitiu à Humanidade adquirir, desde tempos imemoriais, a noção de \textit{indeformável}, 
que está na base da Geometria Clássica, que surgiu assim como ciência experimental muito antes de 
Euclides lhe dar uma formulação matemática.\\

Em Física Clássica nada nos impede de conceber a existência dos corpos \textit{rígido-limites}, isto 
é, dos corpos \textit{o mais rígidos possível}, como \textit{indeformáveis}.
Em todos os tratados de Mecânica Clássica há, assim, um capítulo sobre a Cinemática e um outro sobre
 a Dinâmica destes corpos \textit{rígidos/indeformáveis}, usualmente designados simplesmente 
por \textit{rígidos}.\footnote{Os resultados destes capítulos são condizentes com a observação dos corpos reais. 
 Assim, os 6 graus de liberdade dos corpos rígido/indeformáveis estudados na Cinemática Clássica
 correspondem às translações e rotações que podemos impor aos corpos reais que nos envolvem e nos parecem indeformáveis.}

As palavras \textit{rígido} e \textit{indeformável} são, assim, usadas em Física Clássica como sinónimos,
 embora a primeira corresponda a uma noção puramente geométrica e cinemática e a segunda a uma noção física, 
que não pode ser definida sem referência a forças.\\

Em 1909, Born ~\cite{Born}, usando noções puramente cinemáticas, definiu o corpo \textit{indeformável relativista},
 mas, prolongando em Relatividade a equivalência da Física Clássica, cometeu o erro de lhe 
chamar \textit{rígido}.\footnote{Born, fundamentalmente, definiu a \textit{métrica própria} de um meio contínuo
 em movimento e considerou, em seguida, que os \textit{movimentos indeformáveis} são aqueles em que a métrica 
própria se mantém a mesma. Os \textit{corpos indeformáveis} são, então, os que só podem ter \textit{movimentos indeformáveis}. 
(Os corpos deformáveis podem, obviamente, ter movimentos indeformáveis).}\\

O \textit{rigid body} que encontramos em todos os tratados de Relatividade é, de facto, o 
\textit{corpo indeformável relativista} de Born de 1909.\\

Este erro de designação deu origem a inúmeros paradoxos sobre os quais se escreveram centenas e
 centenas de artigos no século passado e que ainda não apareceram esclarecidos em nenhum 
tratado.\footnote{Os cálculos mostram que a \textit{métrica própria} de um disco a rodar 
num referencial de inércia é não euclideana. Um disco \textit{indeformável} a rodar não pode 
portanto parar porque a sua métrica própria passaria de não euclideana a euclideana.
Como explicar este impedimento de que se não vê sinal na observação dos corpos reais ?
É este o célebre paradoxo do \textit{"disco a rodar"}, também designado numa abordagem mais geral 
por paradoxo dos \textit{"3 graus de liberdade dos corpos rígidos (\textit{na terminologia de Born}) em Relatividade"} ou,
 ainda, por \textit{"paradoxo de Ehrenfest"} que o assinalou em primeiro lugar.}\\

A questão pode, no entanto, ser explicada  num curso elementar de Relatividade. 
 É o que vamos mostrar nas linhas que se seguem.

\vspace{.7 cm}

Em meados do século XVIII Alembert, usando a lei de Hooke, escreveu a sua célebre equação com que mostrou que as ondas
 elásticas se propagam ao longo de uma barra homogénea de densidade $\rho_0 $ e módulo de elasticidade $E$ com uma velocidade :

\begin{equation}
 V =  \sqrt{ \frac{E}{\rho_0} }
\end{equation}

Quando uma barra homogénea $AB$, de comprimento $l_0$, em movimento com uma velocidade longitudinal $v$, choca contra 
uma parede, a extremidade da frente $B$ pára e propaga-se ao longo da barra uma onda de choque a separar a zona parada e 
comprimida da zona ainda em movimento. 

Esta onda atinge a extremidade $A$ no instante : $ t = \Delta t = \frac{l_0}{V}$, contado a partir do instante inicial do choque.
 No instante seguinte, a extremidade $A$ volta para trás com a velocidade $- v$ e, ao longo da barra, propaga-se uma nova onda 
em sentido contrário, a separar a zona de novo em movimento da zona ainda parada e comprimida, que atinge $B$ no 
instante $t = 2 \,\Delta t$, em que toda barra está de novo não comprimida e em movimento.\footnote{A utilização directa 
do principio da conservação da energia, ou do princípio da conservação da quantidade de movimento, permitem chegar a estes
 resultados sem necessidade de utilizar a equação de Alembert. São  bons exercícios de Mecânica básica.}

No caso de uma barra \textit{rígido/indeformável} temos $ E = \infty $, a velocidade da onda de choque é infinita e a barra 
pára instantaneamente.

\vspace{.7 cm}

E em Relatividade? O que é que se passa? \\

Devemos admitir que as ondas de choque se propagam com a velocidade limite $c$ no caso dos corpos \textit{o mais rígidos possível}.\\ 

No caso da barra com o comprimento $l_0$ e a velocidade longitudinal $v$, vemos, tendo em conta a contracção de Lorentz, que a
 barra fica toda parada no instante :

\begin{equation}
\Delta t= \frac{l_0 \, \sqrt{ 1-\beta^2}}{c+v}
\hspace{1.5 cm}\mbox{com} \hspace{1.5 cm} \beta = \frac{v}{c} \,
\end{equation}						
e que, neste instante, o seu comprimento próprio é :

\begin{equation}
l = c \, \Delta t = \frac{c \, l_0 \, \sqrt{ 1-\beta^2} }{ c+ v} < l_0
\end{equation}

Os corpos \textit {mais rígidos possível}, ou, mais simplesmente, os corpos \textit {rígidos relativistas}, são assim corpos
 que se deformam. 

\vspace{.7 cm}

Podemos encontrar as leis elásticas destes corpos usando as fórmulas elementares da Relatividade e um dos dois princípios
 fundamentais de conservação. 

Assim, sendo :

\begin{equation}
  P = \frac{m_0 \, v}{\sqrt{ 1-\beta^2}} =
 \frac {\rho_0^0 \, S \, l_o \, v}{\sqrt{ 1-\beta^2}}
\end{equation}
a quantidade de movimento inicial da barra não deformada (com comprimento $l_0$, secção S e densidade própria {$\rho_0^0$}) e
 velocidade $v$, e $p$ a pressão que a extremidade $B$ exerce contra a parede (ou contra uma barra igual vinda em sentido contrário) 
no intervalo $ [ 0, 2 \, \Delta t ]$, podemos escrever :

\begin{equation} 
 \Delta P = - 2 \, P = - p \, S \, 2 \, \Delta t = 
- \frac{ 2 \, p \,S \,l_0 \, c \, \sqrt{ 1-\beta^2}}{c + v }
\end{equation}
donde, escrevendo : 

\begin{equation}
s = \frac{l}{l_0} ; \,\, s = \sqrt{\frac{1-\beta}{1+\beta}}
\end{equation}
obtemos a lei elástica :

\begin{equation} 
p = \frac{ \rho_0^0 c^2}{2} ( \frac{1}{s^2} - 1 )
\end{equation}

Usando a conservação da energia em vez da conservação da quantidade de movimento, chegamos ao mesmo resultado.\footnote{Esta 
fórmula continua válida no caso das tracções, em que temos $s > 1$, em que, em vez do choque da extremidade $B$, podemos considerar 
que a extremidade $A$ foi bruscamente retida. Uma barra \textit{rígida} pode, assim, ser esticada ilimitadamente sem que a sua 
tensão interna ultrapasse $ p = - \rho_0^0 c^2 / 2 $. Esta particularidade permite responder a uma pergunta curiosa: o que é que
 sucede quando um pescador que tem um peixe preso na extremidade de uma linha o deixa entrar num buraco negro e depois começa a
 recolher a linha? A linha, mesmo no caso limite de ser um \textit{"fio rígido"}, esticará ilimitadamente e o pescador, enquanto
 tiver força para isso, poderá continuar a recolhe-la sem nunca conseguir tirar o peixe do buraco negro onde entrou.}\\

Cálculos igualmente simples permitem-nos mostrar que a densidade própria $\rho_0$ da barra comprimida é dado por :

\begin{equation} 
\rho_0 = \frac{\rho_0^0}{2 }( \frac{1}{s^2} + 1 )
\end{equation}

\vspace{.5 cm}

Estas leis elásticas foram descobertas por Mc Crea ~\cite{Crea}, em 1952, e depois, em 1968, pelo autor destas linhas ~\cite{Brotas1},
mas continuam quase desconhecidas e não se encontram referidas em nenhum tratado.\\

As suas generalizações a mais de uma dimensão permitem, no entanto, esclarecer de um modo claro todos os paradoxos relacionados 
com os corpos rígidos em Relatividade \footnote{ Em Relatividade, um disco indeformável a rodar não pode, de facto, parar, mas um 
disco deformável, como é o caso do disco rígido (no sentido o mais rígido possível) pode, deformando-se, como se deforma, de facto, 
uma moeda quando para depois de ter estado a rodar.

Uma moeda parada num referencial de inércia e a mesma moeda a rodar, é a mesma moeda, mas \textit{em condições físicas diferentes}. Em
 Física Clássica, os sólidos, deformáveis ou não, têm \textit{em iguais condições físicas} a mesma forma, e são 3, e não 6, os seus 
graus de liberdade (ou seja, são 3 os parâmetros que caracterizam os seus diferentes movimentos
 possiveis \textit{nas mesmas condições físicas}).
 Em Relatividade, a situação é exactamente a mesma.} \, \footnote{Os paradoxos referidos resultam unicamente da confusão entre a noção 
geométrica e cinemática de \textit{indeformável} e a noção de \textit{rígido}. 
Sem fazerem esta distinção há, no entanto, autores, que continuam a tentar resolver o problema do disco
 em Relatividade Generalizada, procurando, à custa das deformações do espaço-tempo, encontrar
 os 6 graus de liberdade que não encontram em Relatividade Restrita.}.\\

O que estas leis elásticas nos vêm fundamentalmente dizer é que o movimento dos corpos rígidos em Relatividade tem de ser tratado 
com equações diferenciais e não com a técnica dos graus de liberdade que, mesmo em Física Clássica, só serve para os 
corpos rígido/indeformáveis, e não para os corpos deformáveis.\\

Usando a conservação a da energia, ou a conservação da quantidade de movimento, podemos escrever a equação :

\begin{equation} 
\frac{\partial^2 X}{{\partial x}^2} - \frac{1}{c^2} \frac{\partial^2 X}{{\partial t}^2} = 0 
\end{equation}
que é a equação do movimento longitudinal de uma barra rígida.\footnote{Esta equação tem invariância de Lorentz enquanto que a 
velha equação de Alembert tem invariância de Galileu. A diferença resulta de nela se usarem como variáveis independentes as
 variáveis do referencial $(x,t)$, ditas de Euler, enquanto que na de Alembert se usam as variáveis $(X,t)$, ditas de Lagrange,
 em que $X$ é uma variável associada aos pontos do corpo em movimento.}\\

Por estranho que pareça, esta equação é praticamente desconhecida. Aplicada ao caso dos sólidos não a encontrei em nenhum
 tratado.\footnote{Nos tratados de Relatividade há habitualmente capítulos sobre os \textit{líquidos relativistas} em que é 
apresentada a fórmula $ p = \rho_0^0 c^2 - \rho_0 c^2$ que encontramos quando reunimos as duas fórmulas do $p$ e 
do $\rho _0$ dos \textit{sólidos rígidos} no caso a uma dimensão. Nestes capítulos são estudadas com equações diferenciais as 
ondas elásticas longitudinais, mas não as ondas transversais, que podem existir no caso dos sólidos, mas não dos líquidos. Não
 deixa de ser curioso aparecerem nos tratados de Relatividade capítulos sobre os líquidos relativistas e não sobre os sólidos.}\\

Podem, no entanto, ser-lhe indicadas várias utilizações:

\vspace{.4 cm}

As suas generalizações a 3 dimensões permitem estudar as ondas transversais nos sólidos rígidos; ~\cite{Bento}

\vspace{.4 cm}

A sua utilização em Relatividade Geral permite estudar o modo como um corpo extenso, (no caso um \textit{"fio rígido"}) pode
 atravessar o horizonte de Schwarzschild; ~\cite{Brotas2}

\vspace{.4 cm}

Por último, a sua extensão aos casos não adiabáticos (com transmissão do calor), em que há que considerar um sistema de 
equações ~\cite{Brotas3}, mostra, claramente, porque têm falhado todas as tentativas para encontrar uma variante relativista da
 equação de Fourier com uma única equação: porque a equação de Fourier é, de facto, a equação da transmissão do calor numa
 barra \textit{rígida/indeformável} que não tem sentido considerar em Relatividade. A situação é igual à da Física Clássica em que,
 para estudar a transmissão no calor numa barra deformável é igualmente necessário utilizar um sistema de equações.\footnote{No caso 
adiabático (sem transmissão do calor) a equação da vibração de uma barra pode ser estabelecida por dois caminhos utilizando um dos 
dois princípios fundamentais de conservação. No caso não adiabático (com transmissão do calor) temos de usar mais uma variável,
 a temperatura. Mas, neste caso, os dois caminhos conduzem-nos a equações diferentes e temos assim um sistema de duas equações 
para estudar em conjunto as deformações e a evolução da temperatura.}

\vspace{.4 cm}

Todos estes estudos podiam ter sido feitos antes de 1930 inseridos num programa que, na altura, ficou incompleto: o de encontrar 
uma versão relativista para todas para as equações da Física Clássica. 
Hoje, parece não terem qualquer actualidade.\\

O desconhecimento das equações da Elasticidade relativista não atrasou, com efeito, em nada, o desenvolvimento da Física Teórica
 Moderna nascida da quantificação das partículas e dos campos, mas em que os modelos de sólidos, relativistas,
 ou mesmo clássicos (a equação de Alembert, por exemplo) não desempenham qualquer papel. 
É provável que assim continue a ser durante muito tempo, décadas ou séculos.\\
 
Compreende-se, assim, o desinteresse por estes assuntos da esmagadora maioria dos físicos actuais. \\
 
Não é, no entanto, razão para não ensinar nos cursos elementares a Elasticidade relativista, em que podem ser dados aos 
jovens estudantes problemas inteiramente acessíveis, que contribui para uma melhor compreensão da
 Relatividade \footnote{A palavra \textit{undeformable} quase não é usada na língua inglesa. Assim, quando pedimos a um 
físico inglês para nos traduzir a palavra \textit{"indeformável"} quase invariavelmente ele responde \textit{"rigid"}. A 
definição de \textit{"rigid"} que encontramos nos dicionários é, aliás, a de \textit{"indeformável"}. Para ensinar esta matéria
 em inglês é assim necessário dar um maior uso à palavra \textit{"undeformable"}. Já me informei de que em chinês há dois
 caracteres diferentes para exprimir \textit{"indeformável"} e \textit{"rígido"}.} e que permite  dar resposta
a algumas  perguntas que os melhores estudantes podem fazer.

\hspace{3 cm}

BIBLIOGRAFIA

\end{document}